\documentclass[letterpaper,journal]{IEEEtran}
\usepackage{amsmath,amsfonts}
\usepackage{algorithmic}
\usepackage{algorithm}
\usepackage{array}
\usepackage[caption=false,font=normalsize,labelfont=sf,textfont=sf]{subfig}
\usepackage{textcomp}
\usepackage{stfloats}
\usepackage{url}
\usepackage{verbatim}
\usepackage{graphicx}
\usepackage{cite}
\usepackage[hidelinks]{hyperref}
\usepackage{orcidlink}
\usepackage{diagbox}  
\usepackage{array}    
\usepackage{booktabs} 
\usepackage{textcomp}
\usepackage{tikz}
\usepackage{balance}
\usepackage{pgfplots}
\usepackage{makecell}
\usepgfplotslibrary{groupplots}
\usepackage{pgfplotstable}
\pgfplotsset{compat=1.7}
\usepgfplotslibrary{groupplots}
\usepackage{soul}

\usepackage{xcolor}      
\usepackage{tcolorbox}   

\begin{document}

\title{Autonomous CSI Prediction Framework for O-RAN-Enabled 5G mmWave Vehicular Networks}




\author{IEEE Publication Technology,~\IEEEmembership{Staff,~IEEE,}
\thanks{This paper was produced by the IEEE Publication Technology Group. They are in Piscataway, NJ.}
\thanks{Manuscript received April 19, 2021; revised August 16, 2021.}}

\markboth{Journal of \LaTeX\ Class Files,~Vol.~14, No.~8, August~2021}%
{Shell \MakeLowercase{\textit{et al.}}: A Sample Article Using IEEEtran.cls for IEEE Journals}

\author{\IEEEauthorblockN{Abidemi Orimogunje\textsuperscript{\orcidlink{0000-0002-5119-1578}}, Vukan Ninkovic\textsuperscript{\orcidlink{0000-0002-3187-1314}}, Nemanja Petrovic\textsuperscript{\orcidlink{0000-0003-3268-389X}}, Evariste Twahirwa\textsuperscript{\orcidlink{0000-0001-6758-3659}}, Gaspard Gashema\textsuperscript{\orcidlink{0000-0002-4174-6904}},\\ Lidija Fodor\textsuperscript{\orcidlink{0000-0002-8199-7767}}, Dejan Vukobratovic\textsuperscript{\orcidlink{0000-0002-5305-8420}}}}


\maketitle

\begin{abstract}

Establishing and maintaining 5G mmWave vehicular connectivity poses a challenge due to high user mobility, requiring the design of robust and efficient beam switching procedures. Unlike reactive beam switching based on channel state information (CSI) feedback received from vehicular users, proactive beam switching exploits CSI prediction to prepare in advance for upcoming beam switching decisions. In this paper, we develop a framework for \emph{autonomous} and \emph{self-trainable} CSI prediction for mmWave vehicular users. In the proposed framework, base stations (gNBs) collect and label data sets to train a CSI prediction model both independently and using federated learning (FL). The data set combines data extracted from the CSI feedback and cellular vehicle-to-everything (C-V2X) cooperative awareness messages (CAMs) of surrounding vehicles. The framework is placed in the context of machine learning and artificial intelligence (ML/AI)-based Open RAN (O-RAN) applications (rApps and xApps) fed by realistic real-world mobility and CSI data from the DeepMIMO simulator. Detailed evaluation results demonstrate feasibility, accuracy, and flexibility of the proposed CSI prediction framework.

\end{abstract}


\section{Introduction}

\IEEEPARstart{5}G millimeter wave (mmWave) technology relies on beamforming in multi-antenna systems to enable seamless and reliable connectivity between a base station (gNB) and a user equipment (UE)\cite{6736750,6515173,10121037}. The beam (re)selection procedures are based on an accurate estimation of channel state information (CSI) between gNB and UE, ensuring their continuous beam alignment \cite{8458146,5enescu20205g}. Accurate CSI estimation is the basis for other radio access network (RAN) operations such as radio resource, interference and mobility management \cite{5enescu20205g}. However, maintaining accurate mmWave beam alignment is a complex task even for stationary devices, and it becomes more challenging for dynamic 5G mmWave vehicular links \cite{10333289,9944695,8615988}. Improving conventional beam management methods that rely on reactive use of possibly outdated CSI feedback, recent trends pursue predictive approaches where beam management takes advantage of CSI prediction methods \cite{10021887,9804217}. Accurate CSI prediction as input to mmWave beam management improves beam alignment accuracy while increasing robustness and resilience, and minimizing signaling overhead of beam management procedures \cite{9034044,10770331}. 


CSI prediction relies on receiver channel estimation algorithms, where historical CSI data is utilized for the design of CSI prediction models \cite{9165822,14kim2023deep,15yuan2020machine}. Existing research on channel prediction in wireless communication is focused on different scenarios ranging from low and medium mobility\cite{10620622,10833709}, to high mobility scenarios \cite{9944695,8615988,14kim2023deep}. Different techniques such as compressive sensing, least squares method, autoregressive techniques, sum of sinusoids and machine learning (ML) algorithms are used to acquire CSI data and predict future CSI \cite{9000850,1628683,9910958,10480335}. Linear autoregressive methods and sum of sinusoids work well, using previous and current CSI data to predict future channel behavior within the slow fading scenario, but lack the capacity to adapt to more dynamic environments \cite{1628683,9427230}. ML techniques have shown advantages for predicting channel conditions, especially for high mobility scenarios \cite{10480335,9427230,8395053}. In particular, deep learning (DL) techniques based on recurrent neural networks (RNNs) such as those based on long short-term memory (LSTM) or gated recurrent units (GRU), were used to predict the channel information in time-varying channels while minimizing the pilot overhead. Several works proposed predicting channel conditions in dynamic vehicular channel scenarios using secondary CSI data from different frequency bands, often aided by DL techniques \cite{9034044,9044427,10188816,9744452}. Channel predictions in 5G mmWave channel were carried out using historical channel information in the context of beam management in fast moving scenarios \cite{10021887,10478627,10858124}.

The recent evolution of 5G RAN towards Open RAN (O-RAN) and its native support for ML and artificial intelligence (AI) provides a suitable framework for ML-native RAN functions such as channel prediction. Indeed, RAN intelligent controller (RIC) architecture enables an open application ecosystem through standardised O-RAN interfaces allowing third-party applications such as xApps deployed on the Near-real-time (Near-RT) RIC to perform near-real time RAN control and optimization, and rApps application deployed in the Non-real-time (Non-RT) RIC within the Service management and Orchestration (SMO) to drive non-real time optimization via policy, analytics and ML \cite{10552840}.
Hence, O-RAN provides an ideal platform for the development and deployment of CSI prediction models tailored to different mobility scenarios \cite{30oran_specs_2025,10024837}. CSI prediction affects several downstream tasks, including beam management accuracy and reliability, a crucial feature of robust and reliable 5G mmWave connectivity \cite{8458146,5enescu20205g}. Embedding the CSI prediction and beam management functions into the RIC-based O-RAN architecture provides a flexible environment for deployment and future evolution of intelligent RAN, the approach we pursue in this work \cite{10877797}.


In this paper, we design, develop and evaluate a framework for autonomous and self-trainable ML-based CSI prediction for 5G mmWave vehicular users that is integrated into O-RAN architecture. The proposed framework is based on continuous collection and autonomous labeling of acquired data, generating and maintaining a training data set, and triggering the model (re-)training to generate or update the CSI prediction model. The training data set consists of: i) CSI data collected from 5G mmWave vehicular UEs, for example, by exploiting reference symbols (e.g., CSI-RS) in the cell area, and ii) location, speed and acceleration data set collected by overhearing vehicle-to-vehicle (V2V) cooperative awareness messages (CAM) \cite{9471886}. Our assumption is that the gNB is configured with Rel. 16 cellular vehicle-to-everything (C-V2X) sidelink (PC5) interface that overhears shared $5.9$ GHz V2V channel recovering all CAM messages exchanged between surrounding vehicles. Each gNB logs both time-synchronised input data (CAM data) and corresponding labels (CSI data) and uses it for (re-)training of the CSI prediction model, operating in autonomous and self-training mode.

The proposed framework is placed in the O-RAN context, where we propose in detail the ML operations (MLOps) data flow and life-cycle management of the CSI prediction module, discussing possible training and inference deployment options. In our simulations and evaluation the data used for the CSI model training is extracted from modifications of the DeepMIMO simulator that is capable of recreating real-world urban vehicular environment and generating realistic channel and mobility data \cite{34Alkhateeb2019}. While DeepMIMO itself is not O-RAN specific, it is used here strictly as a high-fidelity channel dataset, whereas the proposed CSI prediction pipeline and deployment workflow are fully compatible with O-RAN data collection and xApp-based operation.
 
 Using the acquired data, we train an RNN architecture with LSTM layers and 1D Convolutional Neural Network (1D-CNN) models as computationally efficient and deployment-oriented solutions for CSI prediction, targeting a favorable accuracy–complexity–latency trade-off compared to more complex sequence modeling architectures (such as the Transformer--based temporal sequence-modeling baseline additionally evaluated in this work).
Training and inference are performed assuming two scenarios: i) for each gNB, an independent CSI prediction module is trained on the local gNB data, or ii) a central aggregated CSI prediction model is obtained using federated learning (FL) by jointly aggregating local gNB models. Extensive numerical and computational complexity evaluation demonstrates that the proposed framework is feasible and that the RNN-based CSI prediction model provides accurate CSI prediction across different realistic DeepMIMO-generated 5G mmWave vehicular scenarios.

In the conference version of this paper\footnote{Conference paper: A. Orimogunje, V. Ninkovic, E. Twahirwa, G. Gashema, and D. Vukobratovic, “Autonomous Self-Trained Channel State Prediction Method for mmWave Vehicular Communications,” in \emph{Proc. 29th Eur. Wireless Conf. (EW),} 2024, pp. 72–76.}, we established the initial framework for autonomous CSI prediction in mmWave vehicular communications and presented preliminary evaluation results based on synthetic mobility traces governed by an Finite State Markov Chain (FSMC) vehicular movement model \cite{10925932}. In contrast, this journal version substantially extends the conference work by replacing the simplified mobility abstraction with realistic vehicular mobility traces and DeepMIMO-based channel reconstruction, thereby enabling evaluation under significantly more realistic and dynamically varying vehicular propagation conditions. Furthermore, none of the results from the conference paper are reused herein, and the present work introduces several additional methodological, architectural, and deployment-oriented extensions, focusing on the following main contributions:

\begin{itemize}
\item We develop a detailed O-RAN-native framework for autonomous and self-trainable CSI prediction in high-mobility mmWave vehicular networks (Section III). The proposed framework describes the design, deployment, orchestration, and integration of CSI prediction modules within the O-RAN ecosystem, including xApp/rApp deployment options, MLOps lifecycle management, and associated deployment trade-offs and challenges.

\item We present the design, preprocessing, training, and inference workflow of the proposed CSI prediction module using synchronized CAM and CSI data streams extracted from realistic vehicular mobility traces and DeepMIMO-based channel reconstruction (Section IV).

\item We evaluate and compare multiple deployable sequence-learning architectures, including RNN-, 1D-CNN-, and Transformer-based predictors, under practical near-RT RIC constraints, analyzing the trade-offs between prediction accuracy, temporal modeling capability, computational complexity, and inference latency (Section IV-A).

\item We investigate federated learning (FL)-based CSI prediction model design for distributed and scalable deployment across multiple gNBs, enabling the development of universal CSI prediction models suitable for O-RAN-native vehicular environments (Section IV-B).
\end{itemize}
We start the next section by providing relevant background to this work and introducing the system model.

\section{System Model}

In this section, we lay out the 5G mmWave vehicular communications system model under consideration. The model consists of a 5G new radio (NR) mmWave gNBs, 5G NR user equipments (UEs) embedded within a mobile vehicular platform, and the mmWave channels between them.


\subsection{5G mmWave Base Station Model}
We consider a 5G NR mmWave gNB deployed in an urban environment covering an area of a busy road/street intersection. gNB operates in mmWave band, i.e., frequency range 2 (FR2) at the carrier frequency $f_c$. Its antenna panel is a 2D uniform linear array consisting of $M=M_x \times M_y$ antennas at a distance of $d=\lambda/2$ from each other in both dimensions, where $\lambda$ is the wavelength corresponding to $f_c$. gNB serves vehicular UEs using 5G mmWave beamforming in the downlink channel of bandwidth $B$ containing $K$ OFDM subcarriers, where beam management decisions are made by a downlink Beam Management Module (DL-BMM). We assume DL-BMM exploits periodic CSI prediction inputs it receives from CSI Prediction Module (CSI-PM) for every attached vehicular UE, where CSI-PM provides periodic CSI prediction service configured for each UE. The focus of our work is the design of autonomously trained and deployed CSI-PM module that provides per-UE CSI prediction service in the context of O-RAN xApps/rApps (see Section III for more details). 

The CSI-PM module collects data from two radio interfaces: i) from 5G NR FR2 interface, the CSI feedback obtained using channel reference symbols (e.g., CSI feedback obtained through CSI-RS symbols) is delivered to CSI-PM module along with the exact acquisition timing (e.g., frame and subframe index), and ii) from supplementary 5G NR C-V2X receiver, gNB overhears  the $5.9$~GHz band to collect CAM messages broadcasted periodically (with a typical period of $\Delta \tau = 100$~ms) by all vehicular UEs in the cell, from which it extracts exact timing, location, speed and acceleration of all surrounding UEs. The two data streams are collected, aligned in time and stored as a dataset used for training the CSI prediction model: a core component of CSI-PM. As we describe in detail in Section. III, the process is essentially autonomous, i.e., the data collection and labeling process can be performed automatically where vehicular UE features such as location, speed, acceleration are associated with the corresponding CSI label taking into account their time alignment. Furthermore, Section III details on how both CSI-PM and DL-BMM modules could be implemented as xApps within O-RAN Near-RT RIC \cite{10877797}.   

\subsection{5G mmWave Channel and CSI Acquisition Model} 

We consider a ray-tracing based multipath wireless channel model applied in a given urban environment. Assuming a user is equipped with $N$ antennas, and for notational clarity focusing first on the \(N=1\) case\footnote{The single receive antenna formulation is adopted for notational simplicity and clarity of exposition. Nevertheless, the proposed CSI prediction framework naturally generalizes to multi-antenna user configurations, where increasing the number of receive antennas primarily increases the CSI output dimensionality. An extended scalability evaluation for \(N=\{1,2,4,8\}\) user receive antennas is presented in Section IV-A and Table~\ref{tab:multiantenna_scalability}}., the channel matrix in the frequency domain $\mathbf{H}^{b,u}_{k,t} \in \mathbb{C}^{M \times N}$ is constructed between a gNB $b$ and a user $u$ on subcarrier $k$ at a given time instant $t$. Taking into account $L$ strongest ray-tracing paths, the vector of channel coefficients $\mathbf{h}^{b,u}_{k,t} \in \mathbb{C}^{M \times 1}$ is:
\begin{equation}\label{eq1}
\mathbf{h}_{k,t}^{(b,u)} = \sum_{l=1}^{L} \sqrt{\frac{\rho_l}{K}} e^{j\left(\vartheta_l + \frac{2\pi k}{K} \tau_l B + 2\pi \nu_l t \right)} \mathbf{a}\left(\varphi_{\text{az}}^{(b,u)}, \varphi_{\text{el}}^{(b,u)}\right)
\end{equation}
where $\rho_l$ and $\vartheta_l$ are the path gain and the phase, $\tau_l$ is the propagation delay, $\nu_l$ is the Doppler shift associated to the signal path $l$ due to user mobility, and $B$ is the bandwidth. The BS array response $\mathbf{a}\left(\varphi_{\text{az}}^{(b,u)}, \varphi_{\text{el}}^{(b,u)}\right)$ is the array vector expressed in Kronecker product form as:

\begin{align}
\mathbf{a}(\varphi_{\text{az}}^{(b,u)}, \varphi_{\text{el}}^{(b,u)}) & =  \mathbf{a}_x(\varphi_{\text{az}}^{(b,u)}, \varphi_{\text{el}}^{(b,u)}) \otimes \nonumber \\
& \otimes \mathbf{a}_y(\varphi_{\text{az}}^{(b,u)}, \varphi_{\text{el}}^{(b,u)}) \otimes \mathbf{a}_z (\varphi_{\text{el}}^{(b,u)}),
\end{align}
where $\mathbf{a}_x(\cdot)$, $\mathbf{a}_y(\cdot)$ and $\mathbf{a}_z(\cdot)$ represent the BS array response vector in $x$, $y$ and $z$ direction, respectively, that changes over time $t$ due to user mobility (see \cite{34Alkhateeb2019} for more details).

In this work, we assume that the values of $\mathbf{h}_{k,t}^{(b,u)}$ are obtained using a ray-tracing tool embedded as part of the DeepMIMO simulator evaluated in a realistic 3D model of urban environment and locations of gNB and UEs \cite{34Alkhateeb2019}. Although the DeepMIMO simulator is not tied to a specific network architecture, it provides high-fidelity, physics-based mmWave channel and mobility datasets widely used for data-driven beamforming and channel prediction studies. Here, it serves solely as a realistic source of channel and mobility data. In contrast, the proposed CSI prediction framework, data processing pipeline, and deployment workflow are developed to operate within the O-RAN architecture, remaining fully independent of the particular dataset generation tool. We consider periodic sampling where the CSI vector $\mathbf{h}_{k,t}^{(b,u)}$ is sampled at regular time intervals $t = n\Delta\tau, n \in \mathbb{N}$. Although in practice a quantised version of CSI is fed back, for simplicity, we assume that gNB receives and collects full CSI information in the form of a matrix $\mathbf{H}^{b,u}_{n\Delta\tau} \in \mathbb{C}^{M \times K}$ that collects channel gains across all $M$ antennas and all $K$ subcarriers.

\subsection{Real-World Vehicular Users Model} 

We consider a fleet of vehicular users moving along the road in both directions. Each vehicle is equipped with 5G NR mmWave (FR2) and 5G NR C-V2X 5.9~GHz radios for vehicle-to-infrastructure (V2I) and V2V communications, respectively. The two interfaces are complementary: the 5.9~GHz C-V2X/PC5 radio provides robust broadcast/control-plane connectivity (e.g., CAM/BSM exchange, neighbor discovery, and coordination/beam-assistance), whereas the FR2/mmWave interface is activated opportunistically for short-range, high-throughput payload transfer \cite{8642796,8887840,9594793}. The former interface provides side information to the gNB about vehicle movements, which is exploited for CSI prediction for the latter interface.

The vehicular mobility model considered in this work follows realistic vehicle movement patterns along a main street with two intersecting streets (see Section IV-A and Fig.~\ref{Figure_4}). The dataset is derived from the Dynamic Doppler scenario of the DeepMIMO framework \cite{36deepmimo_dd1_2024}, which captures realistic high-mobility vehicular propagation conditions and mobility-induced channel dynamics. The considered dataset includes vehicle location (\(x\), \(y\), and \(z\) coordinates), speed, acceleration, and CSI information between moving vehicles and multiple gNBs. Vehicular mobility and CSI data are time-synchronized and sampled with periodicity \(\Delta\tau = 100\) ms across 2000 temporal scenes, where the mobility traces originate from real-world measurements and the corresponding CSI values are generated using the DeepMIMO ray-tracing-based channel reconstruction framework. This setup enables realistic evaluation of the proposed CSI prediction framework under practical vehicular mobility and mmWave propagation conditions.

\section{O-RAN Framework for Autonomous and Self-Trainable CSI Prediction}

In this section, we describe how the proposed autonomous and self-trainable CSI prediction framework can be deployed as part of the 5G O-RAN network architecture.

\begin{figure*}[!t]
    \centering
    \includegraphics[width=\textwidth]{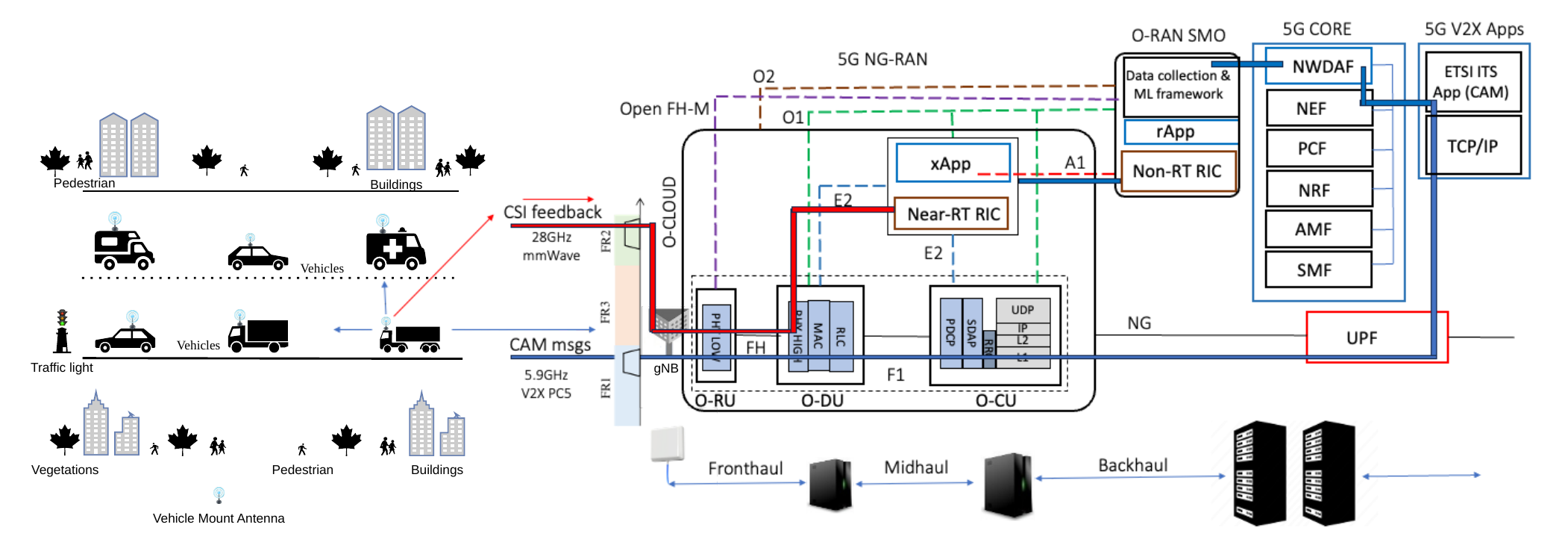}
    \caption{Autonomous
and self-trainable CSI prediction framework deployment options within the 5G O-RAN network architecture.}
    \label{Figure_1}
\end{figure*}

\textbf{Overview of O-RAN Architecture:} The evolution of recent generations of mobile networks is defined through the evolution of the 3GPP technical standards \cite{37lin20243gpp}. Driven by the need for open, multi-vendor, interoperable, AI-driven, and autonomous RAN networks interconnected by open interfaces, O-RAN Alliance produced a set of technical specifications that augment the 3GPP-defined RAN architecture \cite{10024837,30oran_specs_2025}. O-RAN follows the 5G network division into the 5G core (5GC) network and the 5G next generation radio access network (5G NG-RAN). O-RAN builds upon a disaggregated gNB architecture introduced in 3GPP that divides the monolithic gNB into a centralised unit (CU) and a distributed unit (DU). In O-RAN, gNB is divided into centralized unit (O-CU), distributed unit (O-DU) and radio unit (O-RU). In addition to the CU/DU split and the F1 interface defined in 3GPP, O-RAN divides DU into O-DU and O-RU connected via the Open Front Haul (O-FH) interface following the 7.2x functional split. O-RAN components (O-CU/O-DU) can be run as micro-services in containerised environments (e.g., Docker) on general purpose compute resources, making the deployment of O-RAN functions simple, flexible and scalable.

O-RAN augments disaggregated gNB with novel control functions called RAN RIC. Non-RT RIC is hosted as a part of SMO framework and handles control loops at longer time scales ($>1$s). It hosts control applications called rApps that may implement intelligent policies, manage network slices, and provide recommendations based on data analytics. Near-RT RIC handles near-RT control routines ($10$ms-$1$s) and hosts xApps that control and optimize RAN functions within O-DU/O-CU. The O-RAN architecture is illustrated in Fig. \ref{Figure_1}.  

\textbf{ML/AI support in 3GPP/O-RAN:} Both 3GPP and O-RAN are actively working on integration of ML/AI framework in both 5GC and NG-RAN \cite{10353004}. In 5GC, starting from Rel. 15, 3GPP introduced a Network Data Analytics Function (NWDAF) for data collection, model training and inference to support data analytics based on data sources provided by other 5GC network functions (NFs). Integration of ML/AI in NG-RAN is an ongoing 3GPP effort. O-RAN considers Non-RT and Near-RT RICs as locations for ML/AI framework deployment. Both RICs provide support for third party applications: rApps (Non-RT RIC) and xApps (Near-RT RIC), that exploit ML/AI frameworks deployed in respective RICs to collect data, train and deploy AI models (MLOps functionality). In this section, we consider different options for deployment of the proposed autonomous CSI prediction framework set on top of the 3GPP/O-RAN ML/AI framework support. 

\textbf{Collection of Vehicular Mobility and CSI data:} The proposed CSI prediction framework relies on two streams of data: i) vehicle mobility data, ii) CSI data. 

For vehicular mobility data, we assume the gNB collects (overhears) CAM messages from surrounding vehicles using a separate V2X slice that pre-configures Rel. 16 C-V2X sidelink (PC5) interface at 5.9 GHz band at selected gNBs. Note that CAM messages represent third-party application-level data (ETSI ITS application data) that could be collected by a Vehicular Application Server (VAS) and made available to 5GC through the 5GC Network Exposure Function (NEF). In other words, we assume that VAS collects CAM messages from the set of assigned gNBs, extracts relevant mobility data, and push it to 5GC NEF API (e.g., RESTful interface). NEF authenticates and authorises the application (VAS) and forwards this data to Unified Data Repository (UDR): a 5GC network data repository or routes it directly to relevant NF (e.g., NWDAF), as illustrated in Fig. \ref{Figure_1} (blue line data flow).
From NWDAF, either output NWDAF analytics or raw mobility data can be ingested to O-RAN SMO and delivered to data collection and ML framework. Here, mobility data can be used for model training or can be forwarded from Non-RT RIC via A1 interface to Near-RT RIC where it may be consumed for inference by relevant xApps (e.g., CSI-PM module). 

CSI data are collected from the 5G NG-RAN physical layer (PHY) that collects and processes CSI feedback from all active UEs in the cell. Exposing this data in O-RAN architecture is part of ongoing O-RAN standardisation. Certain open-source framework offer access to CSI data at gNB, for example, in OpenAirInterface (OAI), one may exploit T$\_$Tracer tool to extract CSI data of each connected UE on a slot-by-slot basis \cite{10925986}. After being extracted at O-DU, the CSI data could be exposed to other O-RAN network functions (red line data flow). For example, CSI data published by O-DU could be consumed for training and inference by ML/AI frameworks in Near-RT RIC or directly by xApps (e.g., CSI-PM module) via E2 interface using E2 Application Protocol (E2AP). Further, CSI data can be exposed for data logging and training procedures at the Non-RT RIC via A1 interface.  

\textbf{Model Training, Deployment and Inference:} Time stamped vehicular mobility and CSI data are stored in data lakes and used in (re)training the model for CSI-PM module. The standard MLOps framework is an integral part of NWDAF, Non-RT RIC and Near-RT RIC. Thus, data collection and CSI-PM model training could be executed at any of the three locations (the details of CSI-PM model training are provided in Section IV). Once trained, CSI-PM model can be deployed at NWDAF, Non-RT RIC (as an integral part of CSI-PM rApp) and Near-RT RIC (as an integral part of CSI-PM xApp). 

The goal of the CSI-PM framework is to produce CSI prediction every $\Delta \tau$, where $\Delta \tau$ is in the range of $10-100$ms. Following this time dynamics, we propose the CSI-PM inference engine which is deployed as an xApp within Near-RT RIC. We note that in order to provide timely inputs to the CSI-PM inference model, latest vehicular mobility and CSI data samples need to be made available to CSI-PM model at least within time span of a few tens of milliseconds. The exact timing/delay analysis of vehicular mobility and CSI data flows is out of the scope of this paper as it clearly depends on O-RAN deployment configuration (see more details in \cite{10287312}). However, exposure of latest CSI data from O-DU to Near-RT RIC within strict time limits seems to be less challenging than more elaborate ``loop'' undergone by CAM messages (compare red vs blue path in Fig. \ref{Figure_1}). In order to reduce the delay of CAM messages, deployment of distributed (break-out) user plane functions (UPFs) along with relevant 5GC NFs (such as NEF and NWDAF) close to the edge in close proximity of target O-RAN gNBs is a recommended solution. 

\textbf{Outputs and Downstream Processing:} CSI-PM module produces and exposes CSI prediction for any vehicular UE for which this service is active. Any downstream xApp/rApp or NF can consume this data. For example, CSI prediction can be relevant side information for DL-BMM module that could be implemented as another xApp residing in the Near-RT RIC. We leave full integration and analysis of CSI-PM and DL-BMM O-RAN xApp chain for our future work.



\section{CSI Prediction Model: Design, Analysis and Evaluation}

In this section, we present the details of the CSI-PM model design and evaluation under realistic high-mobility vehicular communication conditions. The proposed CSI-PM framework is developed using synchronized vehicular mobility and CSI data extracted from the Dynamic Doppler (DD1) scenario of the DeepMIMO framework (see Section II-C)~\cite{36deepmimo_dd1_2024}, designed to capture realistic mobility-induced mmWave channel dynamics in urban vehicular communication environments. First, we consider CSI-PM model design tailored to a specific gNB, where a separate CSI prediction model is trained and deployed independently for each gNB. Next, we extend the framework toward federated learning (FL)-based CSI-PM design, enabling the development of network-wide CSI prediction models deployable across multiple gNBs while preserving prediction accuracy close to that of locally optimized per-gNB models. For both approaches, we describe the processes of data collection and labeling, model architecture and training, inference workflow, and performance evaluation in terms of prediction accuracy, computational complexity, and deployment-related metrics.

\subsection{CSI-PM Model Design: Real--World Mobility Data}

\textbf{Data collection and labeling:} 
The dataset consists of 5000 periodically sampled temporal scenes containing vehicle location (\(x,y,z\) coordinates), speed, and acceleration recorded with periodicity \(\Delta\tau=100\) ms. These CAM-derived mobility features are used as input to the predictive model, while the corresponding time-aligned complex-valued CSI serves as the prediction target (label). Since the considered scenario includes height variability across vehicles and communication links, the \(z\)-coordinate is retained as an input feature to capture 3D spatial effects on CSI evolution. CSI values are obtained from a 5G NR FR2/mmWave communication setup with a \(4 \times 4\) BS antenna configuration, resulting in \(M=16\) complex CSI coefficients per sample for the \(N=1\) UE antenna configuration considered in the baseline setup. The corresponding system parameters are summarized in Table~\ref{tab:example}. CSI measurements are collected simultaneously from four neighboring gNBs (BS1-BS4), enabling both per-gNB CSI prediction analysis and FL-based distributed model design.

\begin{figure}[!t]
    \centering
    \includegraphics[width=3.4in]{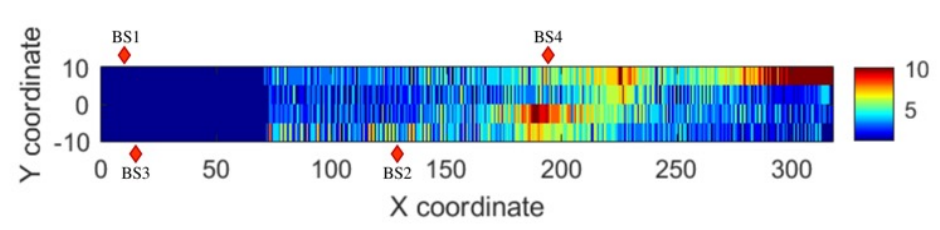}
    \caption{Vehicle density and base station locations in real-world mobility data DeepMIMO scenario.}
    \label{Figure_4}
\end{figure}

The considered vehicular scenario consists of a main street intersected by two secondary streets, enabling realistic variations in user trajectories, mobility states, propagation conditions, and Doppler dynamics. The street layout, gNB locations, and heatmap of vehicle positions illustrating the spatial density of vehicular users relative to the four gNBs are shown in Fig.~\ref{Figure_4}.

\begin{table}[htbp]
    \caption{DeepMIMO gNB Parameters}
    \label{tab:example}
    \centering
    \begin{tabular}{|c|c|}
        \hline
        DeepMIMO parameter & Value \\
        \hline
        Number of gNB antennas $M$ & $16 (4 \times 4)$ \\
        Antenna spacing $d$ & $\lambda/2$  \\
        System bandwidth $B$ & $100$~MHz \\
        No. of OFDM subcarriers $K$ & 240 \\
        Carrier Frequency $f_c$ & $28$~GHz \\
        Number of Ray-Tracing Paths $l$ & $5$ \\
        \hline
    \end{tabular}
\end{table}

In our previous conference work \cite{10925932}, we investigated whether incorporating historical CSI estimates alongside CAM data could improve prediction performance. Although minor gains were observed, the use of previous CSI values introduced additional complexity and potential error propagation. Therefore, in this work we focus exclusively on CAM-derived mobility features, namely vehicle position, speed, and acceleration, as input to the prediction model. This design choice simplifies the system architecture, reduces signaling and computational overhead, and improves robustness in scenarios where prior CSI estimates may be noisy, outdated, or unavailable.

Another challenge arises from the large dynamic range of CSI coefficients, which may span several orders of magnitude depending on the UE-gNB distance and propagation conditions. To ensure stable training and numerically consistent model behavior, each complex CSI coefficient is decomposed into its real and imaginary components, forming a 32-dimensional real-valued target vector for the baseline \(N=1\) configuration. The resulting CSI vectors are normalized using min-max normalization:
\begin{equation} \mathbf{h}_{\text{norm}} = \frac{\mathbf{h} - \min(\mathbf{h})}{\max(\mathbf{h}) - \min(\mathbf{h})}, \end{equation} 
where \(\mathbf{h}\) denotes the concatenated real and imaginary parts of the complex CSI coefficients obtained from the channel model in Section II-B and Eq. (\ref{eq1}). This normalization maps CSI vectors into the \([0,1]\) interval and improves training stability across different propagation conditions and gNB locations.

During preprocessing, invalid or incomplete temporal samples (e.g., synchronization mismatch or missing mobility/CSI entries) are represented as NaN values and excluded through sliding-window validation. Consequently, incomplete or unreliable temporal segments do not propagate into the training or inference pipeline.

\begin{figure}[t]
\centering
\includegraphics[angle=270, width=1\linewidth]{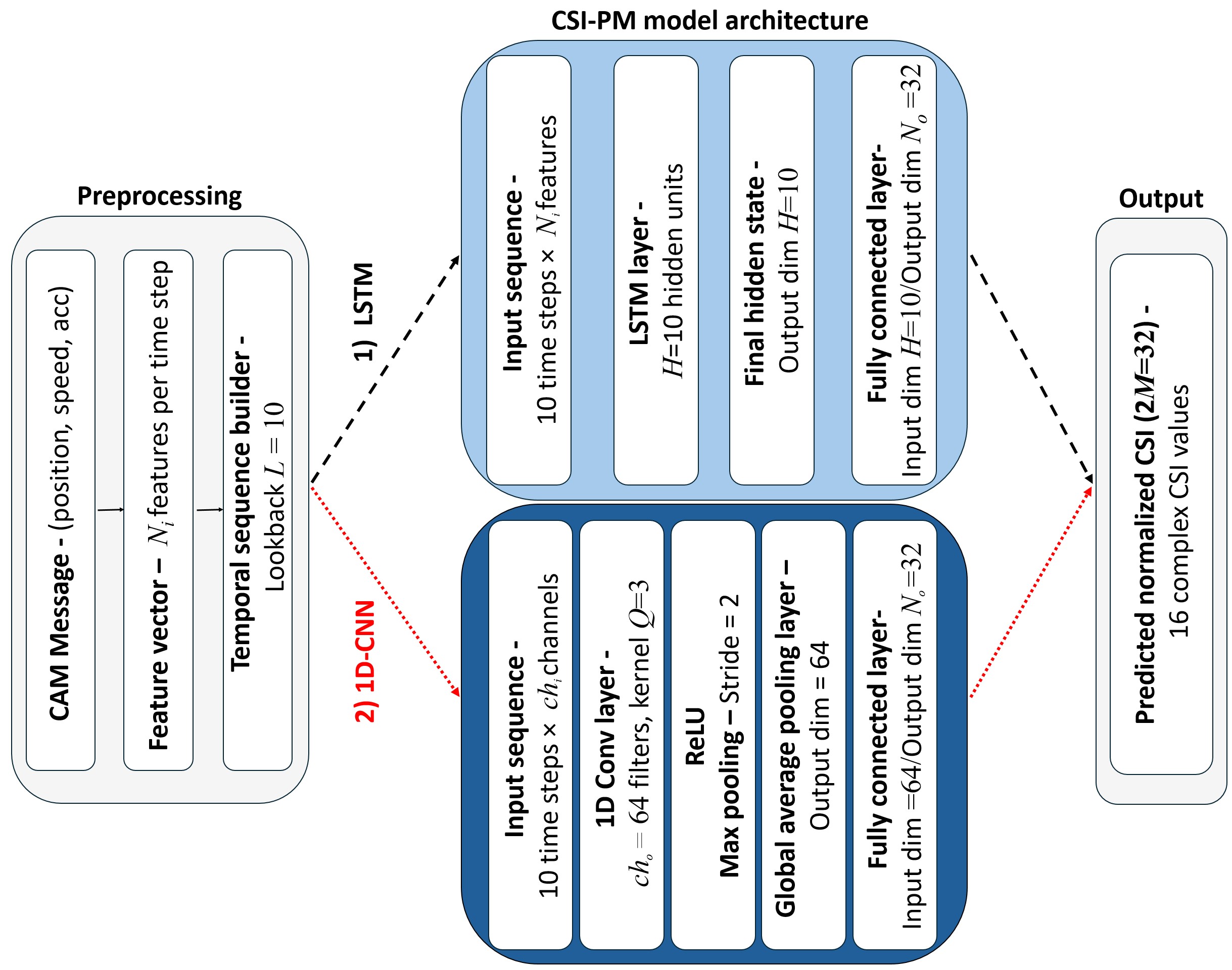}
\caption{Block diagram of the CSI-PM pipeline and model architectures: 1) Proposed LSTM-based CSI-PM predictor (core architecture, black dashed path); 2) 1D-CNN-based architecture (benchmark, red dashed path).}
\label{fig:csi_pm_arch}
\end{figure}

\textbf{CSI-PM model architecture and training:} 
The CSI-PM framework is based on a data-centric predictive architecture that leverages synchronized vehicular mobility and CSI datasets described above. Vehicle position (\(x,y,z\) coordinates), speed, and acceleration extracted from CAM messages are used as input features, while normalized CSI vectors serve as prediction targets. Four independent CSI-PM models, one for each gNB (BS1-BS4), are trained using the corresponding locally collected datasets, enabling specialization to the unique propagation and mobility characteristics observed at each gNB location. The consistent structure of the datasets, including their formatting, normalization scheme, and temporal resolution, directly shapes the design and training pipeline of the CSI-PM models. A block diagram of the CSI-PM pipeline and the evaluated model architectures is shown in Fig.~\ref{fig:csi_pm_arch}, illustrating the preprocessing stage, temporal sequence construction, and the internal layer structure of both the lightweight LSTM-based predictor and the 1D-CNN benchmark.

 The core CSI-PM architecture consists of a lightweight RNN based on a single LSTM layer. The input at each time step is represented by an \(N_i=5\)-dimensional feature vector containing vehicle position ($x$, $y$ and $z$ coordinates), speed, and acceleration information, processed over a temporal sequence of length \(L=10\). The LSTM layer contains \(H=10\) hidden units responsible for capturing temporal dependencies associated with vehicular mobility and CSI evolution. The output of the LSTM layer, with dimension $H$ is passed through a fully connected (FC) layer with \(N_o=2M=32\) ($M = 16$ is the number of complex CSI coefficients) output neurons corresponding to the real and imaginary components of the predicted CSI vector for the \(N=1\) UE antenna configuration (as depicted by black dashed path in Fig. \ref{fig:csi_pm_arch}). This enables the model to reconstruct the future CSI vector based on past motion patterns.

As a lightweight convolutional baseline, we implement a 1D-CNN using a single convolutional block (red dashed path in Fig.~\ref{fig:csi_pm_arch}). The input to the network consists of a temporal sequence with \(ch_i = 5\) channels and a length of \(L = 10\). A convolutional layer with \(ch_o = 64\) filters, each of size \(Q = 3\) and a stride of 1, is applied to this input. This operation produces an output sequence of length \(F = L - Q + 1 = 8\), where each of the 64 filters slides over the input to compute a dot product across each local temporal region. Following the convolution, a ReLU activation function is applied, and a max-pooling layer with a kernel size and stride of 2 reduces the temporal dimension to \(F = 4\). A Global Average Pooling (GAP) layer is then used to average each channel across the remaining temporal dimension, resulting in a fixed-size vector of length \(ch_o = 64\). This compact representation is subsequently passed through a fully connected layer with \(N_o = 32\) output units to generate the final CSI prediction corresponding to the baseline single-antenna (\(N=1\)) UE configuration. The structure is designed to provide a lightweight and computationally efficient benchmark architecture for CSI prediction under near-RT deployment constraints.

To further extend the comparative analysis, we additionally implement a Transformer-based predictor as a modern temporal sequence-modeling baseline using the same input structure (\(L=10\), \(N_i=5\)), preprocessing pipeline, and train/test splits as the LSTM- and 1D-CNN-based CSI-PM models. The Transformer architecture is intentionally configured in a compact form to satisfy lightweight deployment constraints relevant to near-RT RIC operation and edge deployment. Specifically, the model uses an input embedding dimension \(d_{model}=32\), two encoder layers with 4-head self-attention, and a feed-forward dimension \(ff_{dim}=64\), followed by a linear output layer producing the 32-dimensional CSI prediction vector for the \(N=1\) antenna configuration.

To enable systematic comparison of computational complexity between the evaluated architectures, Table~\ref{table:layer_ops} presents approximate FLOP expressions for FC, 1D-CNN, and LSTM layers, while Table~\ref{table:transformer_flops} summarizes the dominant FLOP expressions associated with Transformer encoder operations. Based on these analytical expressions, the resulting approximate FLOP counts for the evaluated CSI-PM architectures are presented in Table~\ref{tab:flops_comparison_real_world}. The obtained complexity analysis highlights the different computational characteristics of the evaluated architectures. The 1D-CNN model is included as a lightweight convolutional baseline suitable for efficient local temporal feature extraction, while the Transformer-based predictor serves as a modern sequence-modeling baseline capable of capturing long-range temporal dependencies through self-attention mechanisms. The proposed LSTM-based CSI-PM architecture provides a balanced recurrent modeling approach with efficient temporal-state representation while also achieving the lowest computational complexity among the evaluated architectures. This makes the LSTM-based CSI-PM particularly suitable for near-RT RIC and edge deployment scenarios with strict latency and resource constraints.

\begin{table}[tbp]
\caption{Approximate FLOPs Counts for FC, 1D-CNN, and LSTM Layers}
\centering
\resizebox{\columnwidth}{!}{
\begin{tabular}{|c|c|c|}
\hline
\diagbox{Layer}{Operation} & Multiplications & Additions \\
\hline
FC & $N_i^{FC} \cdot N_o$ & $(N_i^{FC} - 1) \cdot N_o$ \\
\hline
1D-CNN & $F \cdot Q \cdot ch_i \cdot ch_o$ & $F \cdot (Q \cdot ch_i - 1) \cdot ch_o$ \\
\hline
LSTM & $4 \cdot L \cdot (N_i + H) \cdot H$ & $4 \cdot L \cdot (N_i + H - 1) \cdot H$ \\
\hline
\end{tabular}
}
\label{table:layer_ops}
\end{table}

\begin{table}[tbp]
\caption{Approximate FLOPs counts for a Transformer encoder layer (dominant operations only).}
\resizebox{\columnwidth}{!}{
\centering
\begin{tabular}{|c|c|c|}
\hline
\diagbox{Component}{Operation} & Multiplications & Additions \\
\hline
Q projection  & $L\cdot d_{model}^2$ & $L\cdot d_{model}\cdot (d_{model}-1)$ \\
\hline
K projection  & $L\cdot d_{model}^2$ & $L\cdot d_{model}\cdot (d_{model}-1)$ \\
\hline
V projection  & $L\cdot d_{model}^2$ & $L\cdot d_{model}\cdot (d_{model}-1)$ \\
\hline
Attention scores ($QK^T$) 
& $L^2\cdot d_{model}$ 
& $L^2\cdot (d_{model}-1)$ \\
\hline
Attention output ($\text{Attn}\cdot V$) 
& $L^2\cdot d_{model}$ 
& $L^2\cdot (d_{model}-1)$ \\
\hline
Output projection 
& $L\cdot d_{model}^2$ 
& $L\cdot d_{model}\cdot (d_{model}-1)$ \\
\hline
FFN layer 1 ($d_{model}\!\rightarrow\! ff_{dim}$) 
& $L\cdot d_{model} \cdot ff_{dim}$ 
& $L\cdot ff_{dim}\cdot (d_{model}-1)$ \\
\hline
FFN layer 2 ($ff_{dim}\!\rightarrow\! d_{model}$) 
& $L\cdot ff_{dim} \cdot d_{model}$ 
& $L\cdot d_{model}\cdot (ff_{dim}-1)$ \\
\hline
\end{tabular}}
\label{table:transformer_flops}
\end{table}

All evaluated models are trained to predict the CSI vector at the next temporal observation using a sliding window consisting of the previous 10 CAM snapshots. Due to the sequential nature of the prediction process, the first 10 temporal samples are excluded from training. The LSTM model processes the input as a temporal sequence of shape \(10 \times 5\), whereas the CNN interprets the same input as a multichannel 1D signal of shape \(5 \times 10\). The Transformer-based model processes the input sequence using self-attention mechanisms that capture temporal dependencies across all observation intervals simultaneously.

\begin{table}[htbp] 
\centering 
\caption{Approximate FLOP Counts - Real--World Mobility Setup} \begin{tabular}{|c|c|c|c|} 
\hline 
Architecture & Input Size & Multiplications & Additions \\ 
\hline LSTM + FC & 5 features, 10 timesteps & 6,320 & 5,888 \\
\hline 
1D-CNN + FC & 5 channels, 10 timesteps & 9,728 & 9,184 \\
\hline Transformer &5 features, 10 timesteps&177,984&172,752\\
\hline 
\end{tabular} 
\label{tab:flops_comparison_real_world} 
\end{table}

To ensure scale-invariant learning and enable model comparison across datasets with varying UE-gNB distances, training is performed using the Normalized Mean Squared Error (NMSE) loss function:
\begin{equation}\label{eq4}
    \text{NMSE} = \frac{\mathbb{E}\left[\|\hat{\mathbf{h}} - \mathbf{h}\|_2^2\right]}{\mathbb{E}\left[\|\mathbf{h}\|_2^2\right]},
\end{equation}
where $\hat{\mathbf{h}}$ and $\mathbf{h}$ represent the predicted and ground-truth CSI vectors, respectively. Additionally, training is conducted on 70\% of the samples from each of the four datasets using the Adam optimizer, with a learning rate of $\alpha = 0.00005$, momentum parameters $\beta_1 = 0.9$ and $\beta_2 = 0.99$, and a mini-batch size of 64. This configuration ensures stable convergence while effectively handling variations across normalized CSI samples and mobility conditions. Finally, while training separate CSI-PM models for each gNB enables better adaptation to local propagation environments and mobility patterns, it also introduces scalability and model-management challenges that are addressed in the next subsection through the proposed federated learning (FL)-based framework.

\textbf{CSI-PM model inference and performance evaluation:} Once the training phase is completed, the CSI-PM model is deployed for real-time inference. The model receives input features extracted from each UE’s CAM message and outputs the predicted CSI vector corresponding to the next temporal observation interval. 

We evaluate the CSI-PM model inference on the remaining 30\% samples from the perspective of four different base stations: BS1-BS4, that consume real-world mobility data sets. 
Table~\ref{NMSE_rw_tab} summarizes the NMSE results obtained using the evaluated LSTM-, 1D-CNN-, and Transformer-based CSI-PM architectures.
The obtained results demonstrate that both the LSTM- and Transformer-based predictors consistently outperform the lightweight 1D-CNN baseline across all considered gNB datasets. This behavior confirms the importance of explicitly modeling temporal dependencies in high-mobility CSI prediction scenarios. While the 1D-CNN architecture efficiently captures short-term local temporal patterns through convolutional filtering, both recurrent processing and self-attention mechanisms provide improved capability for modeling mobility-induced channel evolution and long-range temporal dependencies.

Fig.~\ref{Fig_LSTM_1DCNN} illustrates a random test sample from the BS1 dataset, where the LSTM-based CSI-PM model delivers accurate and consistent CSI predictions.  It consistently outperforms the 1D-CNN,
owing to its ability to better capture sequential dependencies
and temporal correlations within the CAM feature stream. The observed performance advantage of the LSTM-based architecture is directly related to the temporal evolution properties of the considered mmWave vehicular channel model in Eq.~\eqref{eq1}. In particular, the Doppler-dependent term \(e^{j2\pi \nu_l t}\) introduces continuous time-varying channel evolution governed by vehicular mobility, resulting in temporally correlated and non-stationary CSI sequences affected by channel aging and multipath evolution. Although CAM features provide instantaneous mobility information (position, speed, and acceleration), the CSI at a given observation interval also depends on the preceding mobility trajectory and accumulated temporal channel evolution across multiple previous intervals. Consequently, CSI prediction in the considered high-mobility scenario becomes a sequential state-evolution problem requiring temporal dependency modeling across consecutive observations~\cite{chen_2023,li_2021}.

Recurrent architectures such as LSTM are particularly suitable for this setting because their internal memory state explicitly propagates temporal information across observation intervals, enabling efficient modeling of Doppler-induced channel evolution and long-range temporal dependencies~\cite{greff_2017, stenhammar_2024}. In contrast, the evaluated lightweight 1D-CNN architecture captures temporal structure through local convolutional filtering over short temporal neighborhoods within the \(L=10\) observation window, but without maintaining an explicit recurrent temporal state. While this enables efficient extraction of short-term temporal patterns, it limits the ability of the 1D-CNN model to capture accumulated mobility-induced CSI evolution and channel aging effects in highly dynamic vehicular environments, leading to slightly higher NMSE values observed in Table~\ref{NMSE_rw_tab} and Fig.~\ref{Fig_LSTM_1DCNN}~\cite{ige_2024}.

The lightweight Transformer-based predictor achieves prediction accuracy comparable to the proposed LSTM-based CSI-PM model across all evaluated scenarios. However, this performance is obtained at substantially higher computational and parametric complexity. Specifically, the Transformer configuration contains 18,208 trainable parameters, compared to 2,142 parameters in the lightweight LSTM model. FLOP-level analysis based on the expressions summarized in Tables~\ref{table:layer_ops} and \ref{table:transformer_flops} indicates approximately \(1.78\times10^5\) multiplications and \(1.73\times10^5\) additions per inference sample for the Transformer architecture, compared to 6,320 multiplications and 5,888 additions for the LSTM-based CSI-PM model (Table~\ref{tab:flops_comparison_real_world}). Similarly, measured inference latency under identical hardware conditions increases from approximately \(11.4\,\mu s\) and \(11.8\,\mu s\) for the LSTM and 1D-CNN models, respectively, to approximately \(102\,\mu s\) for the Transformer-based predictor\footnote{Experiments were conducted on a laptop with the following specifications: Intel Core i7-8565U CPU @ 1.99 GHz, 16 GB RAM, integrated NVIDIA GeForce MX230 (2 GB) and Intel UHD Graphics 620 (128 MB).}.

While larger Transformer configurations could potentially further improve prediction accuracy, such scaling would significantly increase the number of trainable parameters, FLOPs, and inference latency. This would conflict with the lightweight and low-latency design objectives targeted in this work for near-RT RIC and edge deployment in vehicular O-RAN scenarios. Consequently, the Transformer-based predictor is included as a strong modern sequence-modeling baseline, while the proposed lightweight LSTM-based CSI-PM architecture remains more favorable in terms of the overall accuracy--complexity--latency trade-off.

\begin{table}[ht]
    \centering
    \caption{NMSE Performance of LSTM, 1D-CNN and Transformer--based Architectures Across Real--World Mobility Datasets}
    \label{NMSE_rw_tab}
\begin{tabular}{|c|c|c|c|c|}
  \hline
  \diagbox{Architecture}{Dataset} & BS 1 & BS 2 & BS 3 & BS 4 \\
  \hline
  LSTM & 0.0094 & 0.011 & 0.0095 & 0.018\\
  \hline
  1D-CNN &  0.011 & 0.019 & 0.012 & 0.023 \\
  \hline Transformer &0.0096&0.0108&0.0094&0.018\\
  \hline
\end{tabular}
\end{table}

\begin{figure*}[t]
\centering
\begin{tikzpicture}
  \begin{groupplot}[
    group style={
      group size=2 by 2,
      horizontal sep=1.7cm,
      vertical sep=1.7cm,
    },
    width=7cm,
    height=6cm,
    grid=both,
    xlabel={Re},
    ylabel={Im},
    legend style={
      at={(1,0)}, 
      anchor=north,
      legend columns=-1,
      /tikz/every even column/.append style={column sep=0.5cm}
    },
  ]

  \nextgroupplot[title={BS1}]
  \addplot[blue, only marks, mark=o] 
    table [x={x}, y={y}] {./Figs./Fig_4_BS/BS1_true};
  \addplot[red, only marks, mark=x] 
    table [x={x}, y={y}] {./Figs./Fig_4_BS/BS1_predictions};
    
  \nextgroupplot[title={BS2}]
  \addplot[blue, only marks, mark=o] 
    table [x={x}, y={y}] {./Figs./Fig_4_BS/BS2_true};
  
  \addplot[red, only marks, mark=x] 
    table [x={x}, y={y}] {./Figs./Fig_4_BS/BS2_predictions};

  \nextgroupplot[title={BS3}]
  \addplot[blue, only marks, mark=o] 
    table [x={x}, y={y}] {./Figs./Fig_4_BS/BS3_true};
  
  \addplot[red, only marks, mark=x] 
    table [x={x}, y={y}] {./Figs./Fig_4_BS/BS3_predictions};

  \nextgroupplot[title={BS4}, , legend to name=UniversalLegend]
  \addplot[blue, only marks, mark=o] 
    table [x={x}, y={y}] {./Figs./Fig_4_BS/BS4_true};
  \addlegendentry{True CSI values}
  \addplot[red, only marks, mark=x] 
    table [x={x}, y={y}] {./Figs./Fig_4_BS/BS4_predictions};
\addlegendentry{Predicted CSI values}
  \end{groupplot}
\end{tikzpicture}
\begin{center}
\ref{UniversalLegend}
\end{center}


\caption{Comparison of true and predicted CSI values for the best-performing samples: BS1 (top right), BS2 (top left), BS3 (bottom right) and BS4 (bottom left).}
\label{Fig_best_samples}
\end{figure*}
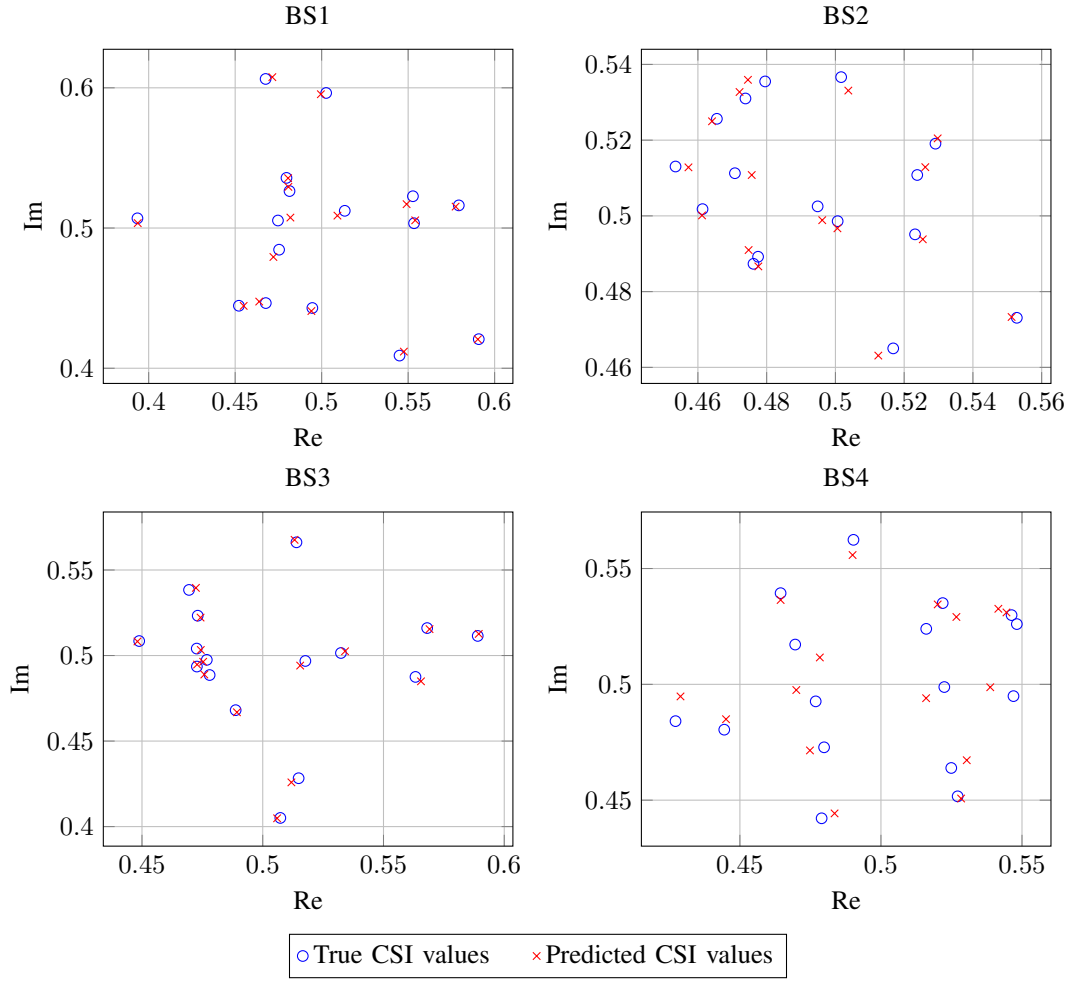

A closer analysis of performance across base stations reveals noticeable differences between the evaluated gNB locations. Specifically, BS2 and BS4, located further from the primary vehicular trajectories (see Fig.~\ref{Figure_4}), exhibit slightly higher NMSE values than BS1 and BS3. This behavior suggests that increased UE--gNB distance and more irregular mobility patterns introduce more complex channel dynamics, ultimately reducing CSI predictability. 
To further illustrate these findings, Fig.~\ref{Fig_best_samples} presents representative predicted CSI samples versus the corresponding ground-truth values across all four BSs. For each BS, the displayed sample is selected by minimizing both the mean squared error (MSE) and the mean absolute error (MAE), with both metrics consistently identifying the same sample. Predictions for BS1 and BS3 closely follow the ground truth, while predictions for BS2 and BS4 exhibit slightly larger deviations but still maintain good overall prediction quality.

\begin{figure}[t]
	\begin{tikzpicture}
  	\begin{axis}[width=0.95\columnwidth, height=6.5cm, 
	legend style={at={(0.75,0.99)}, anchor=north,font=\scriptsize, legend style={nodes={scale=0.95, transform shape}}},
   	legend cell align={left},
	legend columns=1,   	 
   	x tick label style={/pgf/number format/.cd,fixed,
   	 precision=2, /tikz/.cd},
   	y tick label style={/pgf/number format/.cd,fixed, precision=2, /tikz/.cd},
   	xlabel={Re},
   	ylabel={Im},
   	label style={font=\footnotesize},
   	grid=both,   	
   	xmin =0.4, xmax = 0.65,
   	ymin=0.38, ymax=0.6,
   	line width=0.8pt,
   	tick label style={font=\footnotesize}]
    \addplot[blue, only marks, mark=o] 
   	table [x={x}, y={y}] {./Figs./Fig_LSTM_1DCNN/true.txt}; 
   	\addlegendentry{True CSI values}  \addplot[red, only marks, mark=x] 
   	table [x={x}, y={y}] {./Figs./Fig_LSTM_1DCNN/predisctions_LSTM.txt}; 
   	\addlegendentry{Predicted CSI values - LSTM}
    \addplot[black, only marks, mark=square] 
   	table [x={x}, y={y}]
    {./Figs./Fig_LSTM_1DCNN/predictions_1DCNN.txt}; 
   	\addlegendentry{Predicted CSI values - 1DCNN}
  	\end{axis}
	\end{tikzpicture}
	\caption{Comparison of LSTM and 1D-CNN predictions against true CSI values for a single instance from BS1 Dataset.}
	\label{Fig_LSTM_1DCNN}
\end{figure}
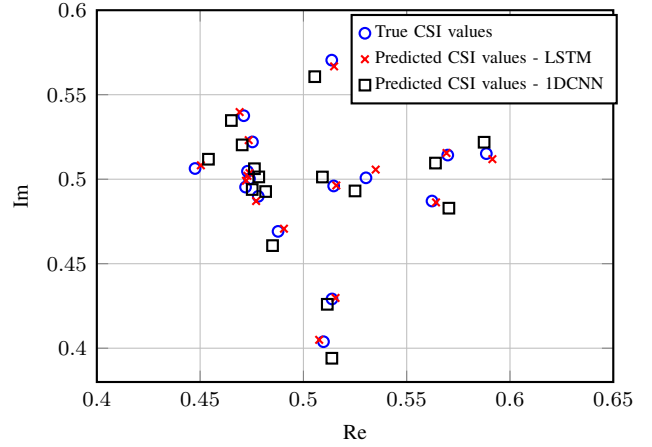

\textbf{Robustness to CSI estimation noise:} In practical deployments, CSI available at the gNB is obtained through pilot-based estimation and is subject to noise and processing errors. To evaluate the robustness of the proposed CSI-PM framework under imperfect CSI conditions, we performed an additional sensitivity analysis by perturbing CSI samples with additive Gaussian noise of controlled power, emulating different CSI estimation quality levels.

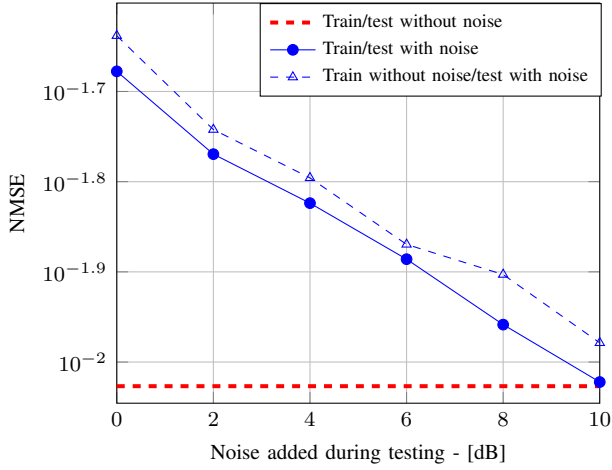
\begin{figure}[t]
\centering
\begin{tikzpicture}
  	\begin{semilogyaxis}[width=0.9\columnwidth, 
	legend style={at={(0.65,1)}, anchor= north,font=\scriptsize, legend style={nodes={scale=1, transform shape}}},
   	legend cell align={left},
	legend columns=1,   	 
   	x tick label style={/pgf/number format/.cd,
   	set thousands separator={},fixed},
   	y tick label style={/pgf/number format/.cd,fixed, precision=2, /tikz/.cd},
   	ylabel={NMSE},
   	label style={font=\footnotesize},   	
   	ymin=0.009,
    ymax=0.025,
    xmin=0,
    xmax=10,
    grid=both,
    xlabel={Noise added during testing - [dB]},
   	tick label style={font=\footnotesize},]
    
   	\addplot[red, dashed, line width=1.5pt] 
    table [x={x}, y={id}] {./Figs./train_test_with_noise_BS1.txt};
    \addlegendentry{Train/test without noise}
\addplot[blue, mark=*] 
    table [x={x}, y={BS1}] {./Figs./train_test_with_noise_BS1.txt};
    \addlegendentry{Train/test with noise}
    \addplot[blue, dashed, mark=triangle, mark options=solid] 
    table [x={x}, y={BS1_ideal_tr}] {./Figs./train_test_with_noise_BS1.txt};
    \addlegendentry{Train without noise/test with noise}

 	\end{semilogyaxis}
	\end{tikzpicture}
\caption{NMSE  versus CSI noise level for BS1 under three scenarios: ideal CSI during training and testing (red dashed curve), identical CSI perturbation during training and testing (solid blue curve), and CSI perturbation applied only during testing (blue dashed curve).}
\label{fig:csi_noise_robustness}
\end{figure}

We consider three scenarios: i) ideal CSI used for both training and testing (results presented in Table~\ref{NMSE_rw_tab}), ii) equally perturbed CSI used during both training and testing, and iii) ideal CSI used for training and perturbed CSI used only during inference. Fig.~\ref{fig:csi_noise_robustness} shows the NMSE performance versus CSI quality for BS1. The results indicate graceful and monotonic performance degradation as CSI quality decreases. For moderate and high CSI quality levels (above approximately 10~dB), prediction accuracy approaches the ideal-CSI benchmark even under mismatch between training and inference conditions. The same qualitative behavior is observed across the remaining BS datasets and is omitted for brevity. These results confirm that the proposed predictor is robust to realistic CSI estimation imperfections.

\textbf{Extended prediction horizon analysis:} To evaluate performance beyond the baseline prediction interval of $\Delta\tau = 100$ms, we conducted an additional study with extended prediction horizons $\Delta\tau = \{100, 200, 300, 400, 500, 800, 1000\}$ms. The same trained CSI-PM configuration and evaluation protocol were used, with the prediction target shifted according to the selected horizon. Table~\ref{tab:delta_tau_bs1} reports NMSE results for a representative BS (BS1). As the prediction horizon increases, a gradual and controlled NMSE degradation is observed, while overall prediction accuracy remains stable even at long horizons. The NMSE increase at $\Delta\tau = 1000$ ms remains limited compared to the 100 ms baseline. The same qualitative behavior and monotonic trend are observed across the remaining BS datasets and are omitted for brevity. These results indicate that the proposed predictor preserves robustness and generalization capability for longer-term CSI forecasting, which is relevant for high-mobility vehicular scenarios and proactive radio resource management.

\begin{table}[ht]
\centering
\caption{NMSE versus prediction horizon $\Delta\tau$ for BS1 under ideal CSI conditions}
\label{tab:delta_tau_bs1}
\resizebox{\columnwidth}{!}{
\begin{tabular}{|c|c|c|c|c|c|c|c|}
\hline
Prediction horizon $\Delta\tau$ [ms] & 100  &200  &300  &400  &500  &800  &1000   \\
\hline
NMSE (BS1)&0.0094&0.0102&0.0108 &0.011& 0.0113 & 0.0114 & 0.0119\\ 
\hline
\end{tabular}}
\end{table}

\textbf{Scalability with multi-antenna UE configurations:} 
To evaluate scalability toward practical FR2/mmWave vehicular deployments, we additionally extend the proposed CSI-PM framework to multi-antenna UE configurations with \(N=\{1,2,4,8\}\) receive antennas while retaining the same \(4\times4\) BS antenna configuration. Increasing the number of UE antennas proportionally increases the CSI output dimensionality, since each receive antenna contributes an additional set of complex CSI coefficients toward the BS antenna panel. To ensure a fair scalability evaluation, the same CSI-PM architecture, temporal processing pipeline, and training procedure are preserved, while only the final fully connected output layer is resized to match the increased CSI dimensionality, resulting in 32 real-valued CSI outputs per UE antenna.

\begin{table}[htbp]
\centering
\caption{Scalability Analysis of the Proposed CSI-PM Framework for Multi-Antenna UE Configurations}
\label{tab:multiantenna_scalability}
\resizebox{\columnwidth}{!}{
\begin{tabular}{|c|c|c|c|c|c|c|}
\hline
\diagbox{$N$}{Metric} 
& \makecell{NMSE \\BS1} 
& \makecell{NMSE\\ BS2} 
& \makecell{NMSE\\ BS3} 
& \makecell{NMSE\\ BS4} 
& \makecell{Inference\\ Time} 
& No. Parameters \\
\hline
1 & 0.0094 & 0.0110 & 0.0095 & 0.0180 & $11.4\,\mu s$ & 2142 \\
\hline
2 & 0.0095 & 0.0110 & 0.0095 & 0.0186 & $12.3\,\mu s$ & 2494 \\
\hline
4 & 0.0095 & 0.0113 & 0.0096 & 0.0190 & $12.9\,\mu s$ & 3198 \\
\hline
8 & 0.0098 & 0.0115 & 0.0097 & 0.0194 & $13.4\,\mu s$ & 4606 \\
\hline
\end{tabular}
}
\end{table}

Table~\ref{tab:multiantenna_scalability} summarizes the obtained results. The proposed framework demonstrates strong scalability across all evaluated configurations. Specifically, the NMSE degradation remains minor even when scaling from \(N=1\) to \(N=8\) UE antennas. At the same time, inference latency increases only moderately from \(11.4\,\mu s\) to \(13.4\,\mu s\), while the number of trainable parameters remains relatively low compared to typical deep sequence-modeling architectures. These results indicate that the proposed lightweight CSI-PM framework can efficiently generalize toward practical multi-antenna FR2 vehicular communication scenarios while preserving favorable low-latency and lightweight deployment characteristics for near-RT RIC and edge-based O-RAN systems.

\subsection{CSI-PM Model Design Using Federated Learning}

\textbf{Data collection and labeling:} The data set in the FL scenario is the same as in Section IV-A. The overall data set consists of four data sets collected at four gNBs (Fig. \ref{Figure_4}) that independently collected mobility and CSI data of a set of vehicles whose mobility is recorded in real world while the corresponding channels are reconstructed using DeepMIMO. At each gNB, the local data set of 5000 samples is divided into $70 \%$ training and $30 \%$ validation. From the validation portion, $25 \%$ was allocated to a centralized test set used to evaluate the global model (see details below).

\textbf{CSI-PM model architecture and training:} Building on the architectures introduced in Section IV-B, where LSTM- and 1D-CNN-based CSI-PM models were presented using five input features, we evaluate model performance within a FL setting. As the LSTM model consistently demonstrated superior performance in terms of prediction accuracy (lower NMSE) and computational efficiency (lower FLOP count), we adopt it exclusively in this subsection. To ensure consistency and fair comparison across scenarios, the same model configuration and training hyperparameters are retained. 

In the FL setup, each gNB acts as a client in the federated network, training a local instance of the CSI-PM model using only the locally collected dataset. A central FL server (e.g., deployed at 5GC NWDAF or at Non-RT RIC as rApp) is responsible for orchestrating the learning process by aggregating model updates from participating gNBs and distributing the updated global model. This decentralized approach allows each gNB to adapt the CSI-PM model to local channel conditions and mobility patterns without sharing raw CSI data, as only model parameters or gradient updates are exchanged during FL rounds. All measurement data remains local to each gNB and within the mobile network operator domain, under standard operational and security controls. Privacy preservation is therefore a secondary benefit of the chosen architecture, while the primary motivation for FL in this work is scalability, robustness, and deployment flexibility across distributed sites. A detailed analysis of privacy leakage and adversarial robustness in FL-based CSI prediction is outside the scope of this work and is considered as an important direction for future study.

FL enables collaborative training of a global model through multiple communication rounds~\cite{43li2020review,44bharati2022federated}. At the beginning of each round, the server sends the current global model parameters to the gNBs. Each gNB then performs local training using its own dataset and returns the updated model weights to the server. These updates are aggregated using a predefined algorithm to produce a new global model, which is broadcasted back to the gNBs for the next round. This process repeats for a fixed number of rounds or until convergence. In our implementation, we use four FL clients (one per gNB), all active in each round, over 20 communication rounds. Each client performs 10 local training epochs per round. This configuration strikes a balance between model convergence and communication efficiency, enabling effective knowledge sharing across gNBs while maintaining manageable computational demands.

To implement this architecture, we employ the Flower FL framework\cite{45beutel2020flower}, which supports both simulated and real-world FL deployments. On the server side, Flower includes a FL loop and a client manager. The FL loop orchestrates the training process and applies the selected aggregation strategy. In our setup, we use FedAvg \cite{46mcmahan2017communication}, a widely adopted and communication-efficient algorithm that averages model updates from clients to generate the global model. The client manager coordinates participation by the gNBs and handles communication. On the gNB (client) side, local logic handles model parameter reception, training, evaluation on local data, and returning the results to the server. A key advantage of Flower is that the server remains agnostic to the internal workings of each client, enabling interoperability across heterogeneous device types and deployment settings. The FL architecture allows the CSI-PM model to scale across many gNBs while maintaining data privacy guarantees and supporting adaptation to localized wireless environments.

We note that the numerical experiments in this work assume benign and reliable gNB clients without adversarial behavior or severe data outliers. In operational networks, gNBs are operator-managed infrastructure nodes subject to continuous monitoring and fault-management procedures, which reduces the likelihood of persistent undetected anomalous updates. Nevertheless, in scenarios involving malfunctioning or compromised clients, more robust FL strategies, such as personalized FL~\cite{dinh_2020} or Byzantine-resilient aggregation rules~\cite{wu_2023}, can be integrated to mitigate model poisoning and client-drift effects. A systematic evaluation of such robust aggregation mechanisms in the considered CSI prediction setting is left for future work.

\textbf{CSI-PM Model Inference and Performance Evaluation:} To assess the effectiveness of the federated CSI-PM model, we conduct both client-side and server-side evaluations, as supported by the Flower framework.

In client-side evaluation, each gNB evaluates its personalized model on a local validation set before model aggregation. Validation losses are aggregated on the server using a weighted average. Over the course of 20 communication rounds, this metric steadily improves, reaching a final NMSE of 0.0159. This outcome indicates that local models effectively adapt to their respective environments and that the federated process promotes consistent improvement in local inference quality.

In server-side evaluation, a centralized test set is used to assess the performance of the global model after each aggregation step. This test set is constructed by sampling 25\% of the validation data from each client, providing a balanced view of generalization performance. The NMSE of the global model improves from an initial value of 1.1264 to 0.0165 at the end of training. This result shows that the FL model is not only capable of learning meaningful representations across diverse clients but also generalizes well across gNBs characterized by varying channel conditions and mobility profiles.

Although the best performance of individually trained models (Section IV-A) reached an NMSE of 0.0094 (BS1, Table \ref{NMSE_rw_tab}), the federated model performs competitively despite never accessing the raw data. The slight drop in predictive accuracy is a trade-off for the substantial gains in scalability, data privacy, and robustness to heterogeneity across sites. Additionally, the ability to generalize through collaboration is particularly beneficial when client datasets are small or non-representative on their own. These findings validate the FL paradigm as a powerful alternative to centralized training for CSI prediction, particularly in distributed wireless networks where data locality, privacy constraints, and dynamic conditions require decentralized learning.

\section{Conclusions}

In this paper, we proposed an O-RAN-compliant framework for autonomous and self-trainable CSI prediction tailored for 5G mmWave vehicular networks. Addressing the limitations of reactive beam management in high-mobility scenarios, our approach evaluates ML models, specifically, RNN-based architectures with LSTM, 1D-CNNs, and simple Transformer-based models to enable lightweight CSI prediction. The framework supports both standalone and FL modes, allowing flexible deployment of CSI prediction modules at individual gNBs or in an aggregated, collaborative manner. A key novelty lies in combining CSI feedback data with context-rich CAMs overheard via the C-V2X PC5 interface, thereby enriching the training dataset with mobility-aware features.

We placed the proposed system within the O-RAN architectural context, detailing the MLOps lifecycle of data acquisition, labeling, training, and inference. Using data generated by an enhanced DeepMIMO simulator that replicates urban vehicular environments, we demonstrated the feasibility, robustness, and accuracy of the prediction models across realistic high-mobility scenarios. The results confirm that self-training, context-aware CSI prediction significantly improves prediction quality, making it a strong candidate for integration into AI-native 5G—and future 6G—RANs. Future work will explore online learning capabilities, scalability across larger deployments, and integration with downstream beam management and scheduling algorithms within the RIC-based O-RAN ecosystem.

\vspace{7pt}
\section*{Acknowledgment}
This work was jointly supported by the African Center of Excellence in Internet of Things (ACEIoT), College of Science and Technology, University of Rwanda, and The Partnership for Skills in Applied Sciences, Engineering and Technology (PASET) Regional Scholarship and Innovation Fund (RSIF). In addition, this work received funding from the Horizon 2020 research and innovation staff exchange grant agreement No 101086387.

\bibliographystyle{IEEEtran}
\bibliography{IEEEabrv, References}

 

\vspace{11pt}

\vfill

\end{document}